\documentclass[12pt]{article}
\usepackage{verse}
\usepackage[a4paper,text={5.6in,9.1in},centering]{geometry}
\usepackage[greek,english]{babel}
\usepackage[applemac]{inputenc}

\verselinenumbersright

\title{A Solution to Galileo's Enigma ``Mostro Son Io''}
\author{G. Busetto and A. De Angelis\\{\small{Dipartimento di Fisica e Astronomia ``Galileo Galilei'' dell'Universit\`a di Padova}}}

\begin{document}

\maketitle

\abstract{Galileo Galilei was a skilled writer and  explored several genres, from the well-known scientific writings (often in the form of dialogs)  to theater and poetry. His last published poem,  {\em Mostro son io} (A Monster am I), is a riddle written in the form of a sonnet.
We suggest that the solution to Galileo's riddle is the Zodiac.}

\vskip 1 cm

To the second part of {\em La Sfinge}, a collection of riddling sonnets by Antonio Malatesti published in Firenze in 1643 \cite{malatesti2} one year after Galileo's death, a sonnet entitled {\em Enimma} (Enigma, or Riddle) by Galileo himself  was added in the beginning of the book. Malatesti had received the sonnet by Galileo as a sign of appreciation: in the first part of {\em La Sfinge}, published in Venice in 1640 \cite{malatesti}, the fifth sonnet was indeed dedicated to ``Galileo's telescope". {\em Enimma} was published with no solution.
Galileo was a skilled writer and  explored several genres, from scientific writing (often in the form of dialog)  to theater and poetry \cite{favaro,ale} and this is probably his last poem.

The original
text of the {\em Enimma} is reported here:
\begin{verse}
\poemlines{1}
 Mostro son io pi\`u strano e pi\`u diforme\\  \poemlines{5}
Che l'arp\'ia, la sirena o la chimera;\\
N\'e in terra, in aria, in acqua \`e alcuna fiera,\\
Ch'abbia di membra cos\`i  varie forme.\\  \vskip 2mm 
     Parte a parte non ho che sia conforme,\\
pi\`u che s'una sia bianca e l'altra nera;\\
Spesso di cacciator dietro ho una schiera,\\
Che de' miei pi\`e van rintracciando l'orme.\\ \vskip 2mm 
\poemlines{9}
     Nelle tenebre oscure \`e il mio soggiorno;\\  
Ch\'e se dall'ombre al chiaro lume passo,\\
Tosto l'alma da me sen fugge, come\\  
\vskip 2mm  
\poemlines{12}
     Sen fugge il sogno all'apparir del giorno;\\
E le mie membra disunite lasso,\\
E l'esser perdo, con la vita, e 'l nome. 
\end{verse}
that we translate as:
\begin{verse} \poemlines{1}
A monster am I, stranger in shape and form\\  \poemlines{5} 
Than a Harpy, a Siren or a Chimera;\\
Nor has on Earth, in air, in water, any beast\\
Such a varied forms of limbs.\\ \vskip 2mm

No part of mine to another does conform\\ 
More than if one were white and the other black;\\
Often of hunters behind me I have a host,\\
Tracing the footprints of my feet.\\ \vskip 2mm
\poemlines{9}
In the obscure darkness is my sojourn:\\
If from the shadows to  clear light I pass,\\
Quickly my soul from me flees out,\\ \vskip 2mm  \poemlines{12}

Just as a dream flees out at break of day:\\
And  disjoined my limbs  I leave,\\
And my being I lose, my life and name.
\end{verse}

Despite various suggestions had been proposed since then for who could be the ``monster'' of the riddle, there is no agreement on the solution to this day. 

In the introduction to his 
Galileo biography \cite{camerota}, Camerota suggests  that it is Galileo himself, although he finds this solution not completely persuasive. Another possible solution, suggested by Bignami,  is `a comet' \cite{bignami}; however  {\em Enimma}  does not refer
to, or suggest, anything contained in Galileo's extensive writing about comets \cite{peterson}. Recently Peterson suggested the solution to be the
Ptolemaic system of the Universe \cite{peterson}: a very brave proposal, but  unpersuasive to us, and somehow forced. Daniele \cite{daniele} suggests a gambit:
the solution could be the telescope  -- the same solution as for Malatesti's sonnet dedicated to Galilei. Another suggestion, by Bartezzaghi, is \cite{bartezzaghi} that the solution is the enigma itself: also a gambit, facinating but also, to us, questionable.

After a discussion, one of us (G.B.) had an intuition, that was elaborated by the other (A.D.A.), and then together. We present it here: we are convinced that the solution is the Zodiac. 

According to 
the most diffuse ethimological interpetation \cite{cambridge}, the word  Zodiac means ``[circle of] little animals'', and originates from the Greek zoidiak\'os, from zoidion  (`small animal' or `sculptured animal figure', diminutive of zoion, `animal'). The name reflects the prominence of animals (and mythological hybrids) among the twelve signs. A discussion on the etymology is present in a book well known to Galilei: Kepler's {\em De Stella Nova} \cite{stellanova}.
The first 6 verses seem to refer to this. In addition,  the signs of the Zodiac were integral to understanding how the human body functioned, and each one was mapping a different part of the body in medical astrology, a ``science'' having its roots in Greek astrology, though it fully bloomed in the Middle Age. The correspondence, called {\em melothes\'ia}, was discussd by  Marcus Manilius (1st century AD) in his epic poem  {\em Astronomica}, well known to Galilei \cite{sources} and to Kepler \cite{stellanova}.

Verses 7 and 8 seem to refer to the constellation of Orion, the giant hunter. Orion is not on the path of the ecliptic, and is ``behind'' the Zodiac. 
According to a Greek cosmogonical myth reported for example by Gaius Iulius Higinus (64 BC -- 17 AD) in his {\em Poeticon Astronomicon,} Gaia, the goddess of  Earth and protector of the animals, was angered by Orion who was chasing all animals, and asked Scorpio, a giant scorpion, to kill him before he could harm them. Scorpio accomplished his mission and chased Orion on the opposit side of the Zodiac. Notice that the word `cacciator' (`hunter') in the original sonnet, in Italian, is truncated with the effect that it can be singular or plural -- although the word ``schiera'' (host) inclines to the plural. 

The last 6 verses are quite clear. In particular, verse 13 seems to refer to the visual disappearance of the constellations in the morning which causes losing 
track of the images associated to them (the stars with greater megnitude disappear later).

Of course the riddle might refer to the only constellation. Scorpio is a possible candidate. Another is the  Ophiucus, the serpent bearer (on the zodiacal circle but not listed among the Zodiac signs for astrological/hystorical reasons), together with the serpent itself: Orion's stars ``follow'' Ophiucus' feet (verse 8), and Galilei studied in detail the 1604 {\em stella nova} in Ophiucus' left foot \cite{lezioni}. However these hypotheses  appears less immediate. And in any case, in the riddling game,  the concept of
``right answer'' contains some degree of arbitrariness.

\paragraph{Acknowledgement --} We thank our colleague Maria Teresa Musacchio for check and suggestions on the English translation of Galilei's sonnet, and Alessandro Bettini for comments.


\begin{thebibliography}{99}
\bibitem{malatesti2}  {\em La Sfinge, Enimmi del Sig. Antonio
Malatesti, Parte seconda,} Stamperia di Sua Altezza Serenissima, Firenze 1643
\bibitem{malatesti}  {\em La Sfinge, Enimmi del Sig. Antonio
Malatesti,} Sarsina, Venezia 1640
\bibitem{favaro}{\em Le Opere di Galileo Galilei,} Edizione Nazionale, 
Antonio Favaro ed.,  vol. I-XX, Barbera, Firenze 1890--1909 
\bibitem{ale} A. De Angelis, {\em Galileo and a lost poem on the 1604 supernova,} arXiv: 2204.04001, April 2022
\bibitem{camerota} M. Camerota, {\em Galileo Galilei,} Salerno, Roma 2004
\bibitem{bignami} G. Bignami, in {\em G. Galilei, Against the Donning of the Gown; Enigma,} Moon Books, London 2000
\bibitem{peterson}M. Peterson, Academia Letters, Article 2758 (2021)
\bibitem{daniele} A. Daniele, ``Galileo Letterato'', Padova 2009, in {\em Intorno a Galileo,} CLEUP, Padova 2022
\bibitem{bartezzaghi} S. Bartezzaghi, {\em Incontri con la sfinge,} Einaudi, Torino 2004
\bibitem{cambridge} Merriam-Webster Thesaurus, https://www.merriam-webster.com
\bibitem{stellanova}J. Kepler, {\em De stella nova in pede serpentarii,} Prag 1606, Chapter 5
\bibitem{sources} C. Hall,  {\em Galileo's Reading,} Cambridge University Press, 2013
\bibitem{lezioni} G. Galilei, Lecture Notes on the {stella nova} appeared in 1604,  in \cite{favaro}, vol. II


\end{thebibliography}
\end{document}